\documentclass[letter, longauth, lineno]{aa}     

\usepackage{graphicx}
\usepackage{txfonts}
\usepackage{xcolor}
\usepackage{lipsum}
\usepackage{subcaption}         
\usepackage{lscape}             
\usepackage{placeins}           

\usepackage{hyperref}    
\usepackage{longtable}
\usepackage[T1]{fontenc}
\graphicspath{ {Figures/} }  
\defcitealias{2024AA...684A..96O}{OM24}
\defcitealias{2022AA...666A.183K}{K22}

\begin{document}

   \title{A tidally detached super Neptune on a strongly misaligned retrograde orbit}

   \author{G. Mantovan
          \inst{\ref{inst1},\ref{inst2},\ref{inst3}}$^{\orcid{0000-0002-6871-6131}}$, L. Malavolta\inst{\ref{inst2},\ref{inst3}}$^{\orcid{0000-0002-6492-2085}}$, A. F. Lanza\inst{\ref{inst4}}$^{\orcid{0000-0001-5928-7251}}$, F. Marzari\inst{\ref{inst5}}, L. Naponiello\inst{\ref{inst6}}$^{\orcid{0000-0001-9390-0988}}$, K. Biazzo\inst{\ref{inst7}}$^{\orcid{0000-0002-1892-2180}}$, R. Cosentino\inst{\ref{inst8},\ref{inst4}}$^{\orcid{0000-0003-1784-1431}}$, M. C. D'Arpa\inst{\ref{inst9}}$^{\orcid{0009-0004-5914-7274}}$, S. Desidera\inst{\ref{inst2}}$^{\orcid{0000-0001-8613-2589}}$, G. Guilluy\inst{\ref{inst6}}$^{\orcid{0000-0002-1259-2678}}$, D. Nardiello\inst{\ref{inst3},\ref{inst2}}$^{\orcid{0000-0003-1149-3659}}$, A. Sozzetti\inst{\ref{inst6}}$^{\orcid{0000-0002-7504-365X}}$, S. Vissapragada\inst{\ref{inst10}}$^{\orcid{0000-0003-2527-1475}}$, R. Aloisi\inst{\ref{inst11}}$^{\orcid{0000-0003-2822-616X}}$, S. Benatti\inst{\ref{inst8}}$^{\orcid{0000-0002-4638-3495}}$, L. Borsato\inst{\ref{inst2}}$^{\orcid{0000-0003-0066-9268}}$, R. Claudi\inst{\ref{inst2},\ref{inst12}}$^{\orcid{0000-0001-7707-5105}}$, S. Jenkins\inst{\ref{inst13}}$^{\orcid{0000-0001-9827-1463}}$, V. Nascimbeni\inst{\ref{inst2}}$^{\orcid{0000-0001-9770-1214}}$, G. Piotto\inst{\ref{inst3}}$^{\orcid{0000-0002-9937-6387}}$, T. Zingales\inst{\ref{inst3},\ref{inst2}}$^{\orcid{0000-0001-6880-5356}}$,
        }

   \institute{Centro di Ateneo di Studi e Attivit\`a Spaziali ``G. Colombo'' -- Universit\`a degli Studi di Padova, Via Venezia 15, IT-35131, Padova, Italy\label{inst1}; \email{giacomo.mantovan@unipd.it}
            \and
             INAF - Osservatorio Astronomico di Padova, Vicolo dell'Osservatorio 5, IT-35122, Padova, Italy\label{inst2}
             \and
             Dipartimento di Fisica e Astronomia ``Galileo Galilei'', Università di Padova, Vicolo dell'Osservatorio 3, IT-35122, Padova, Italy
              \label{inst3}
             \and
             INAF - Osservatorio Astrofisico di Catania, Via Santa Sofia 78, IT-95123, Catania, Italy\label{inst4}
             \and
             Dipartimento di Fisica e Astronomia, Università di Padova, Via Marzolo 8, IT-35121, Padova,Italy\label{inst5}
             \and
             INAF - Osservatorio Astronomico di Roma, Monte Porzio Catone, IT-00043, Roma, Italy\label{inst7}
             \and
             Fundación Galileo Galilei-INAF, Rambla José Ana Fernandez Pérez 7, ES-38712, Breña Baja, Spain\label{inst8}
             \and
             INAF - Osservatorio Astronomico di Palermo, Piazza del Parlamento 1, IT-90134, Palermo, Italy\label{inst9}
             \and 
             INAF - Osservatorio Astrofisico di Torino, Via Osservatorio 20, IT-10025, Pino Torinese, Italy\label{inst6}
             \and 
             Carnegie Science Observatories, CA 91101, Pasadena, USA\label{inst10}
             \and 
             Department of Astronomy, University of Wisconsin, 475 N. Charter Str., WI 53706, Madison, USA\label{inst11}
             \and
             Dipartimento di Matematica e Fisica, Universit\`a Roma Tre, Via della Vasca Navale 84, IT-00146, Roma, Italy\label{inst12}
             \and
             Department of Physics and Kavli Institute for Astrophysics and Space Research, MIT, MA 02139, Cambridge, USA\label{inst13}
             }

    \date{Compiled: \today}
 
  \abstract 
   {The obliquity between a planet's orbital axis and its host star's spin axis provides crucial insights into planetary formation and migration. Planets with scaled semi-major axes ($a/R_\star$) large enough to be unaffected by tidal alterations (``tidally detached''), offer a unique opportunity to study the original obliquity in which the system formed. We therefore observed TOI-1710 b ($a/R_\star \approx 36$) in-transit using HARPS-N + GIANO-B, collecting high-precision radial velocities to measure the Rossiter-McLaughlin (RM) effect. Spectral analysis of the H$\alpha$ and HeI triple lines was also pursued to evaluate atmospheric photoevaporation. Using our knowledge of the star rotation period ($21.5 \pm 0.2$~d), we estimated a true obliquity of $\psi = 149 ^{+11}_{-10}$ deg, which indicates a retrograde motion and places TOI-1710 b among the most misaligned systems -- and the only one known orbiting a cool star in retrograde motion. The strong misalignment favours a high-eccentricity migration (HEM) origin for this low-density super-Neptune planet in the savanna region, challenging previous findings that claimed a minor role of HEM in this period-radius(-density) domain. Moreover, the strong misalignment and lack of a detected close stellar companion suggests a purely planetary post-migration misalignment, likely due to planet-planet scattering followed by planet-planet Kozai-Lidov oscillations and tidal circularisation.}


    \authorrunning{Giacomo Mantovan et al.}

   \maketitle
   \nolinenumbers

\section{Introduction}
The architecture of planetary systems encodes information about their formation and migration histories. In this context, measuring the orbital obliquity -- the angle between a planet's orbital angular momentum and its host star spin angular momentum -- has proven fundamental. The sky-projected obliquity ($\lambda$) can be detected with in-transit radial velocities (RVs) via the Rossiter-McLaughlin (RM) effect \citep[e.g.][]{2000A&A...359L..13Q}. This can be translated into the 3D orbital obliquity ($\psi$) with respect to the stellar rotation axis thanks to the knowledge of a reliable stellar rotation period, projected rotational velocity, and stellar radius, hence stellar inclination ($i_\star$). Star-planet tidal interactions may alter the obliquity, and close-in planets orbiting a few Gyr old stars with sizeable convective envelopes are expected to show aligned configurations, making the formation mechanism inaccessible. By observing systems with scaled orbital semi-major axes ($a/R_\star$) large enough to avoid tidal alterations to their obliquity, we can access the original obliquity configuration in which the system formed \citep[e.g.][]{2023AJ....166..217W}. We categorise these systems as ``tidally detached'' \citep{2021AJ....162..182R}, where the tidal alignment timescale is longer than the system age \citep[][]{2012MNRAS.423..486L}.

Super-Neptune exoplanets ($4 R_\oplus \lesssim R_{\rm p} \lesssim 8R_\oplus$) are an intriguing population for studies of planetary formation and evolution. While the dearth of short-period ($P \lesssim 3$~d) ones is well-known as the ``Neptune desert'' \citep[e.g.][and references therein]{2016A&A...589A..75M}, only recent studies have shown additional structures on the Neptune planet distribution. Neptunes pile up in a ``ridge'' \citep[$3-6$~d,][]{2024A&A...689A.250C} before falling off into the ``savanna'' \citep[$P \gtrsim 6$~d,][]{2023A&A...669A..63B}. Properties such as bulk densities, eccentricities and host-star metallicities change significantly within the desert, ridge, and savanna, hinting that these sub-populations likely experienced different formation and migration histories \citep[e.g.][]{2020Natur.583...39A, 2023Natur.622..255N, 2025AJ....169..117V}. To validate a complete theory, it is necessary to determine not only the role of different migration pathways, but also that of atmospheric evaporation. It is still unknown whether the transition from quiescent to hydrodynamical escape occurs within the desert or the ridge, or further out into the savanna \citep{2025A&A...701A.190B}.

We present a new study of the $\sim$24-day orbit super-Neptune TOI-1710 b \citep[][K22, OM24]{2022AA...666A.183K, 2024ApJS..272...32P, 2024AA...684A..96O} located in the savanna. The large $a/R_\star$ ($\approx 36$) makes it tidally detached, which means that we can access the original, unaltered obliquity configuration and investigate the processes that led to its current architecture.

\section{Observations and data reduction}
\label{sec:obs}

We observed TOI-1710 ($m_{\rm V}$ = 9.5) with the TNG telescope in the GIARPS observing mode \citep[][and references therein]{2017EPJP..132..364C}, which enable the simultaneous measurement of high-resolution spectra in the optical ($0.39 - 0.69~\mu$m, $R \sim 115~000$, HARPS-N) and the near-infrared ($0.95 - 2.45~\mu$m, $R \sim 50~000$, GIANO-B). We collected in-transit (5.2 hours) and suitable off-transit (2 hours) observations on February 2, 2024, obtaining 42 spectra with $t_{\rm exp} =$~600~s. We used the ESPRESSO Data Reduction Software \citep[DRS,][]{2021A&A...645A..96P} v3.2.0, optimised for HARPS-N \citep[][]{2021A&A...648A.103D}, and computed the RVs using the cross-correlation function (CCF) method \citep{2002Msngr.110....9P} with a G2 mask. We proceeded in this way to extract the RVs (including the literature data), rather than using a template matching pipeline, because our RM model simulates the CCF anomaly caused by the transiting planet.

\textit{TESS} observed TOI-1710 at 2 min cadence in sectors 19, 20, 26, 40, 53, 60, 73 and 79. We extracted lightcurves using the \texttt{PATHOS} approach \citep{2020MNRAS.495.4924N}. \textit{TESS} sectors 73 and 79, previously unpublished, were essential for refining the transit ephemeris of TOI-1710 b (uncertainty of about 1 min at the night of February 2) required for the RM modelling.   

\section{Analysis}
\label{sec:analysis}

\begin{figure}
    \centering
   \includegraphics[width=0.85\hsize]{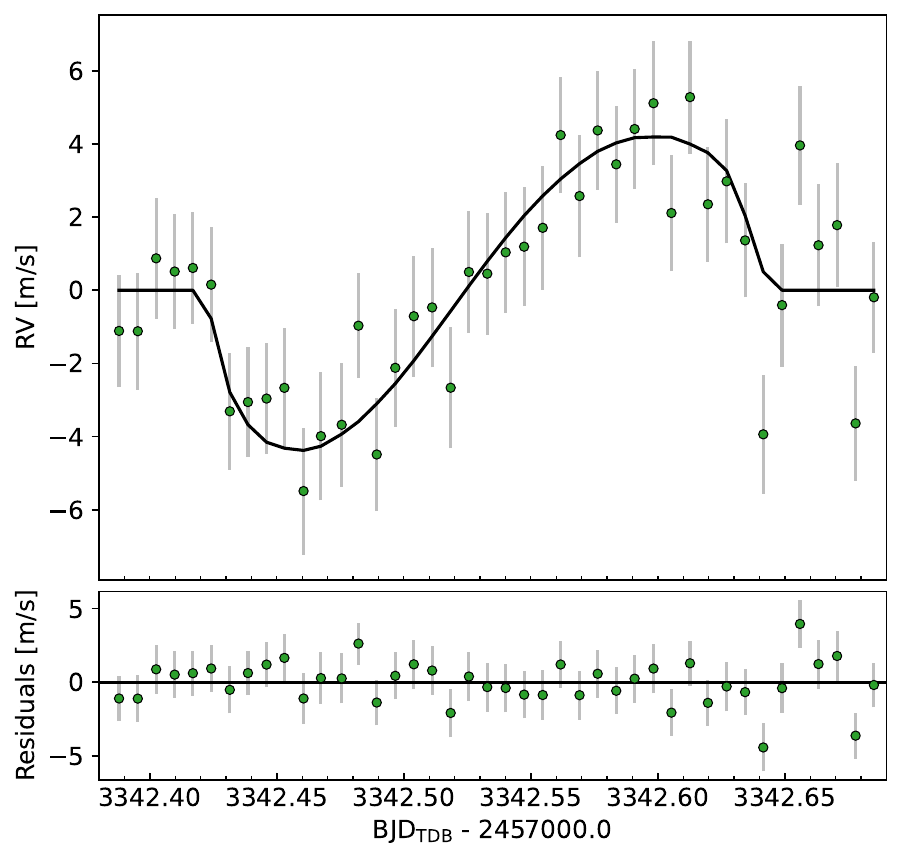}
   \caption{RM fit to the in-transit RV of TOI-1710 b. \textit{Top}: Best-fit (line) over the RVs data (dots), corrected for the Keplerian RV curve and systemic RV. \textit{Bottom}: Residuals of the fit. }
   \label{fig:RM}
\end{figure}

We simultaneously modelled each \textit{TESS} sector with all the HARPS-N off-transit RVs \citepalias[from ][and Sect. \ref{sec:obs}]{2022AA...666A.183K, 2024AA...684A..96O}, and our in-transit RV anomaly. The detailed description of our analysis can be found in Appendix \ref{sec:appB}. The best-fit RM and RV models are shown in Figs. \ref{fig:RM} and \ref{fig:RV}, fitting values in Tables \ref{table:model-rm} and \ref{table:model-rm2}. Additionally, we performed a transmission spectroscopy analysis to search for signs of atmospheric photoevaporation; the results and methodology are summarised in Appendix \ref{app:photo}.

\subsection{Transit ephemeris and planet parameters refinement}
We refined the planetary radius, mass and bulk density, achieving precisions of 2\%, 22\%, and 22\% respectively. The radius value agrees with all previous studies, whereas our mass is 1$\sigma$ lower. This could be attributed to an improved treatment of stellar parameters and activity, as \citetalias{2022AA...666A.183K} did not account for its presence, \citetalias{2024AA...684A..96O} used an incorrect stellar density value in the modelling, and \citet{2024ApJS..272...32P} mention their methodology may provide inaccurate mass estimates for active stars. Such precisions are necessary to constrain the planetary composition, i.e., the planet mass fraction of rock, gas and water, and enable comparative atmospheric studies across the Neptune desert, ridge, and savanna. The low bulk density ($\rho_b = 0.50 \pm 0.11$ g~cm$^{-3}$) of TOI-1710 b is consistent with the trend noted in \citet{2024A&A...691A.233C}, where planets in the savanna typically have $\rho_b < 1$ g~cm$^{-3}$. Based on this threshold, the authors identified a dichotomy in the Neptunian population, separating low-density ``fluffy'' planets, such as TOI-1710~b, from higher density, ``dense'' ones.

\subsection{Rossiter-McLaughlin effect}
From the Bayesian modelling of the RM effect, we obtain a sky-projected obliquity $\lambda\, = 176^{+29}_{-27}$~deg. Then, thanks to the estimation of a reliable stellar rotation period ($P_{\rm rot}\, = 21.5 \pm 0.2$~d, $v \sin{i_\star} = 2.2^{+0.1}_{-0.2}$ km s$^{-1}$, see Appendix. \ref{sec:rot} and $v \sin{i_\star}$ extraction details in \citetalias{2022AA...666A.183K}, where the HARPS-N and SOPHIE spectra indicate $v \sin{i_\star} > 2$ km s$^{-1}$), we translated $\lambda$ into the true 3D orbital obliquity ($\psi$). We did so by sampling from the posterior distributions of $\lambda$, stellar inclination ($i_\star$) and planetary orbital inclination ($i_{\rm p}$), and using Eq. (7) from \citet{2007AJ....133.1828W}: $\cos\psi = \sin{i_\star}\cos{\lambda}\sin{i_{\rm p}} + \cos{i_\star}\cos{i_{\rm p}}$, to estimate $\psi = 149 ^{+11}_{-10}$ deg. This result confirms the retrograde motion and favours a strongly-misaligned orbit (see Fig. \ref{fig:RM}).

\section{Discussion}
\label{sec:discussion}

\subsection{The retrograde orbit of TOI-1710 b}

\begin{figure}
   \centering
   \includegraphics[trim={0.2cm 1.7cm 0.2cm 2.7cm},clip,width=0.85\hsize]%
   {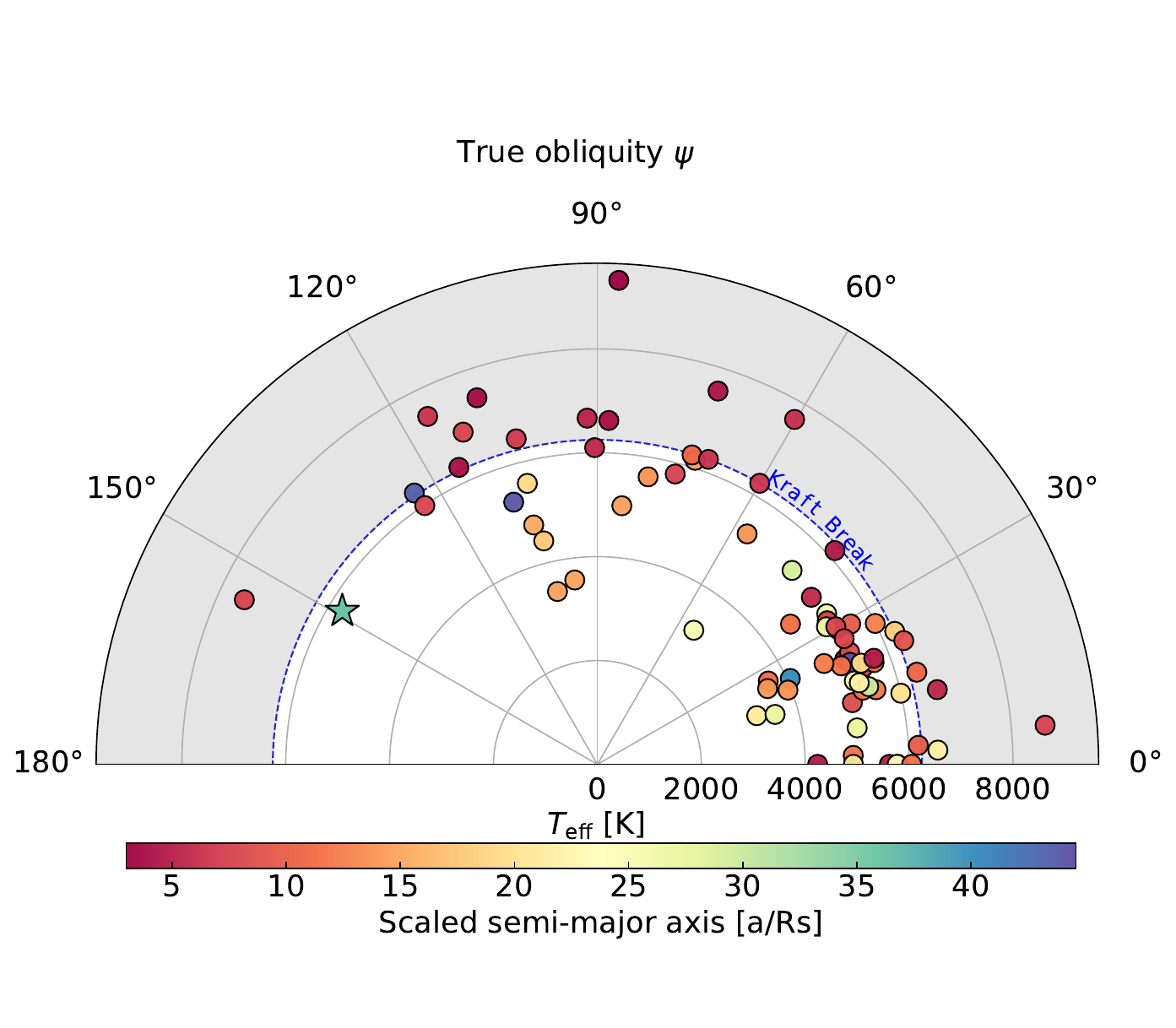}
   \caption{True obliquity of known exoplanets vs. host star effective temperature. TOI-1710~b is marked by a star. The $a/R_\star$ is colour-coded. The shaded region denotes $T_{\rm eff} > 6250$~K. Data: TEPCat and the NASA Exoplanet Archive in February 2026. }
   \label{fig:polar}
\end{figure}

The 3D obliquity of TOI-1710 b stands in the top 1\% of all obliquity measurements when compared alongside those of each known system with a $\psi$ measurement \citep[cf. TEPCat,][]{2011MNRAS.417.2166S} in a polar plot (Fig. \ref{fig:polar}). Moreover, among planets orbiting cool stars \citep[$T_{\rm eff} < 6250$~K, the Kraft break,][]{1967ApJ...150..551K}, TOI-1710 b has the largest $\psi$ -- the only one known indicating a retrograde motion with certainty. TOI-1710 b is located slightly above the group of misaligned systems with polar orbits with $\psi$ in the range 80 -- 125 deg \citep{2021ApJ...916L...1A, 2023A&A...674A.120A}, and supports the trend noted in \citet{2024A&A...684L..17M} where all planets orbiting cool stars in polar (or larger misalignment) orbits have scaled distances $a/R_\star > 10$. This trend may support that only such cool star, tidally--detached systems have preserved their original misalignments, while those with small $a/R_\star$ (orbiting cool stars) have either been tidally realigned or aligned from birth.

\begin{figure}
   \centering
   \includegraphics[trim={0.2cm 0.2cm 0.2cm 0.2cm},clip,width=0.85\hsize]%
   {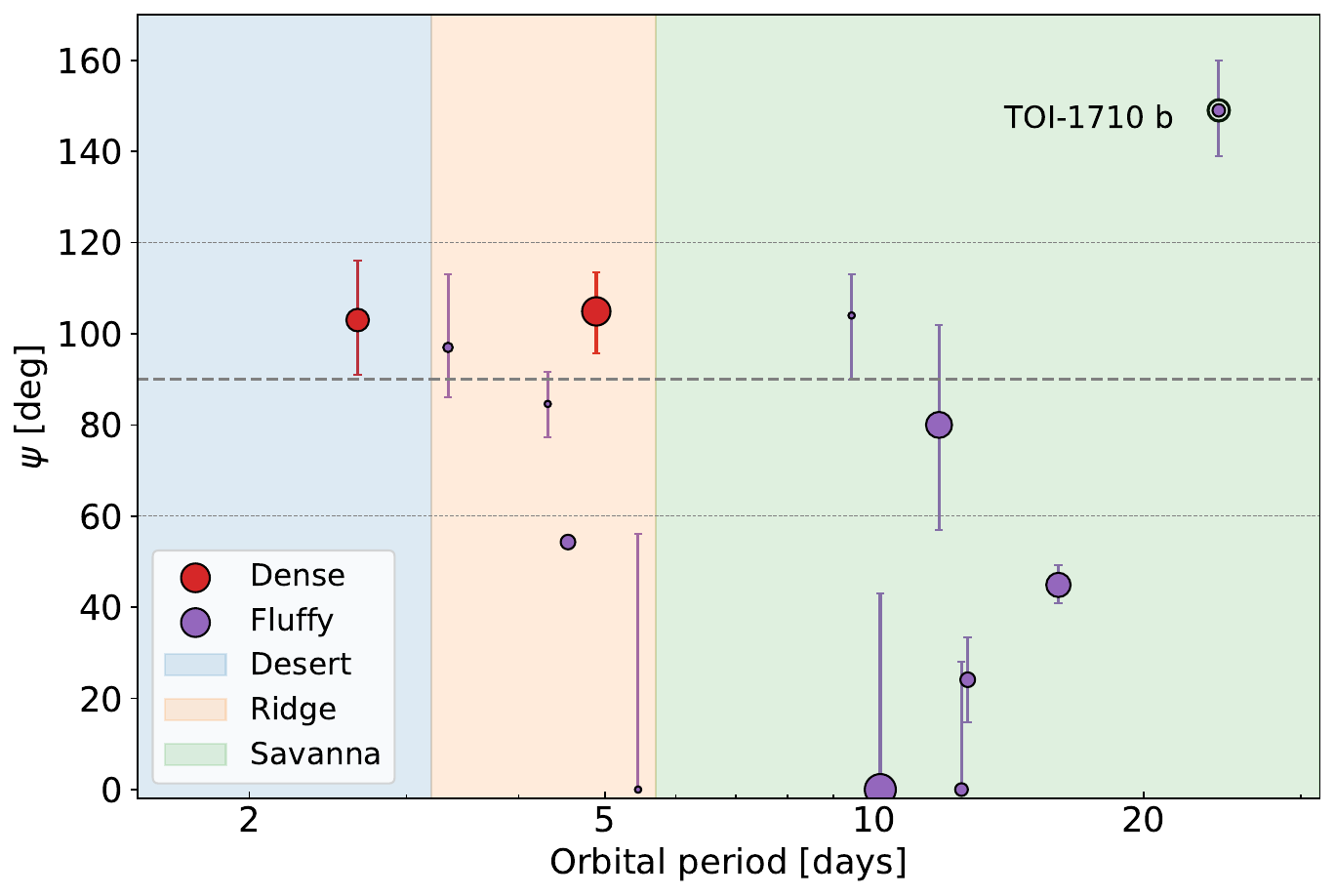}
   \caption{True obliquity of super-Neptunes vs. orbital period. Planetary densities are differentiated by colour, while eccentricities by size. The shaded areas separate the desert, ridge, and savanna. }
   \label{fig:sub-jovians}
\end{figure}

The highly misaligned, retrograde orbit of TOI-1710 b favours a high-eccentricity migration \citep[HEM,][]{rasioford1996, 2003ApJ...589..605W, 2006A&A...453..341M} scenario for this fluffy super-Neptune in the savanna (Fig. \ref{fig:sub-jovians}). This finding seems to disfavour previous hypotheses \citep{2024A&A...691A.233C, 2025A&A...701A.190B} that most fluffy planets -- those in the savanna -- reached their locations via disc-driven migrations \citep{1996Natur.380..606L}, whereas the HEM primarily happened in the case of dense Neptunes, bringing them to the ridge and desert. The characteristics of TOI-1710~b rather suggest that fluffy planets formed beyond the ice line may undergo HEM, possibly triggered by a phase of planet–planet scattering, and reach misaligned orbits without losing their envelopes. This scenario could indicate that HEM may play a role in planetary migration across the entire Neptune period range and over a wider density range. TOI-1710~b is also the longest-period Neptune with a known $\psi$.

\subsection{Wide-binary companion and dynamical implications}
Previous high-resolution imaging studies have revealed no close companions within 1.2$\arcsec$ (and from the diffraction limit of 20 mas) and down to contrast limits of $\Delta \rm mag =$ 5.5 -- 7 \citepalias{2024AA...684A..96O}. This corresponds to physical separations of approximately 1.6 -- 97 au, considering the distance of TOI-1710 from the Earth ($d =$ 81 pc). On the other hand, \citetalias{2024AA...684A..96O} and \citet{2024A&A...689A.302G} confirmed the presence of a wide binary companion located 44.5$\arcsec$ away -- originally reported in \citet{2001AJ....122.3466M}. Using the \textit{Gaia} DR3 parallax \citep{2023A&A...674A...1G}, this angular separation translates to a projected distance of about 3600 au. We followed \citet{2023A&A...677L..15S} and conducted a sensitivity analysis of \textit{Gaia} DR3 astrometry for companions, based on the re-normalised unit weight error statistic \citep[RUWE, ][]{2021A&A...649A...2L}. Such companions would have produced excess residuals in \textit{Gaia} DR3 astrometry, resulting in a RUWE larger than that measured for TOI-1710 (0.94, compatible with a single-star model). We can rule out brown dwarf companions within 8--10 au, while the sensitivity to super-Jovian companions is limited to 3-4 au. However, the latter can be excluded due to the low peak-to-peak RV residuals ($\sim$ 4 m\,s$^{-1}$), which would only translate to sub-Jovian masses at 4 au (see Appx. \ref{sect:det} and  \ref{sec:appC}).

These findings are crucial to place constraints on the formation and dynamical history of TOI-1710 b. Highly misaligned, retrograde orbits typically originate from dynamical mechanisms involving a close stellar companion \citep[e.g.][]{2019Galax...7...84M, 2022PASP..134h2001A}. As a consequence, we can likely rule out primordial misalignment scenarios for TOI-1710 b, since only a protoplanetary disc tilt induced by a close companion \citep[see e.g.][]{2021PNAS..11817418H} would produce the observed $\psi$. It is worth noting that, provided there is a significant mutual inclination between the disc and the binary orbit, even distant stellar companions can misalign the orbital plane of a disc in less than its lifetime  \citep{2012Natur.491..418B}. However, the nodal recession period of the disc scales approximately as $\tau_{\rm disc} \propto a^{\prime~3}/M^{\prime}$, with $a^{\prime}$ being the binary separation and $M^{\prime}$ the companion mass. Therefore, scaling the result of \citet{2012Natur.491..418B} for $a^{\prime} = 500$~au and $M^{\prime}=1 M_\odot$ (i.e. 1.8 Myr) to the parameters of TOI-1710 B ($a^{\prime} = 3600$~au and $M^{\prime} \approx 0.4 M_\odot$) would imply a much longer timescale ($> 500$~Myr) than the disc lifetime.

We therefore investigated post-formation mechanisms. Two dynamical processes (or the interplay between them) could explain the observed retrograde orbit: Kozai--Lidov cycles \citep{2003ApJ...589..605W} and planet--planet scattering \citep{2006A&A...453..341M}. In particular, Kozai-Lidov cycles, especially when the distant companion is on an eccentric orbit \citep{2016ARA&A..54..441N}, have been proposed as the most likely explanations for the retrograde orbits observed in all systems with $\psi$ greater than 120 deg: WASP-131 \citep{2023MNRAS.522.4499D}, K2-290 \citep{2021PNAS..11817418H} and KELT-19 \citep{2024MNRAS.528..270K}, all of which have close stellar companions at separations of 38, 110, and 160 au.

In the case of K2-290, \citet{2022ApJ...925L...5B} discussed the role of a wide binary companion in setting the initial conditions for secular interactions and hence the observed orbital misalignment. However, even in this scenario, the nearby companion at 110 au plays a critical role. Similar to the mechanism proposed for K2-290, \citet{2025ApJ...986..117Y} introduced the ``eccentricity cascade'' non-secular evolution mechanism to explain the retrograde orbit of HAT-P-7 b. However, even this mechanism involving a wide binary requires an intermediate close-orbit stellar companion. For TOI-1710 b, in the absence of such a close companion, we estimated the timescale that Kozai--Lidov oscillations would require if triggered by the wide stellar companion TOI-1710~B alone. We found a timescale on the order of $2 \times 10^{10}$ years, making this scenario highly unlikely. On the other hand, a planet--planet Kozai--Lidov cycle involving a hypothetical giant planet at 30 au would yield a much shorter timescale of about $10^{7}$ years. However, this scenario requires the presence of a second planet in an eccentric, misaligned and relatively nearby orbit (not yet been detected in the RVs, see Sect. \ref{app_twothree}), which could be the result of planet--planet scattering. We note that in its present orbital configuration, the timescale for apsidal precession (driven primarily by general relativity) is of the order of $3.7 \times 10^{5}$~yr \citep{2002ApJ...573..829M}; that is, much shorter than the Kozai--Lidov timescale. Therefore, Kozai--Lidov oscillations could only have played a role in the orbital evolution of TOI-1710 b if it migrated from a more distant location, where the precession timescale would have been longer than the Kozai--Lidov oscillation timescale.

Taken together, these constraints suggest a complex formation history for TOI-1710 b. One possible hypothesis is a multi-step dynamical evolution, with planet--planet scattering \citep[or a stellar fly-by, e.g.][]{2011MNRAS.411..859M} followed by secular Kozai-Lidov cycles, which may have led to inward migration due to tidal interaction with the star. The two planets could have formed at wide separations from the host star, interacted with each other, and the inner one migrated via HEM to arrive at the final orbit in its current misaligned configuration. A notable analogue might be HD80606 b \citep{2001A&A...375L..27N}, a highly eccentric planet in the midst of a HEM, where planet--planet Kozai \citep{2008ApJ...678..498N} is favoured over star--planet Kozai \citep{2009ApJ...703.2091W} due to the wide separation ($\sim$ 1200 au) of its stellar companion. Our findings support the hypothesis that both super-Neptune and single-star systems are primordially aligned and may become misaligned during the post-disk phase \citep[see ][]{2024AJ....168..116R, 2025AJ....169..189R}.

\section{Key findings}
\label{sec:conclusions}

\begin{enumerate}
    \item TOI-1710 b is a tidally--detached planet on a strongly misaligned (top 1\% of all $\psi$ measurements), retrograde orbit -- the only one known orbiting a cool star with accessible $\psi$. It is a ``fluffy'' super-Neptune, and is the longest-period one with a $\psi$ measurement, as well as being the only one with a retrograde orbit. This is a likely signature of HEM;
    \item The HEM mechanism could play a role in the planetary migration of Neptune planets over a wide period (ridge and savanna) and density range. This finding contradicts previous indications that such a mechanism plays only a minor role in shaping the Neptune savanna -- and, more specifically, fluffy ones -- and will help to improve complete migration models;
    \item TOI-1710 has a wide stellar companion but lacks close ones, thus favouring a post-formation origin for the observed misalignment over a primordial one. All other systems with $\psi > 120$ deg have instead a close stellar companion;
    \item While retrograde orbits are typically attributed to star--planet Kozai migration induced by an originally misaligned close stellar companion, our work proposes a distinct, purely planetary formation history. The retrograde orbit of TOI-1710 b is most likely the result of planet--planet scattering, followed by secular planet--planet Kozai-Lidov oscillations and a final orbit circularisation due to tidal dissipation inside the planet. 
\end{enumerate}

\begin{acknowledgements}
Based on observations made with the Italian \textit{Telescopio Nazionale Galileo} (TNG) operated by the \textit{Fundación Galileo Galilei} (FGG) of the \textit{Istituto Nazionale di Astrofisica} (INAF) at the \textit{Observatorio del Roque de los Muchachos} (La Palma, Canary Islands, Spain). This paper includes data collected with the TESS mission, obtained from the MAST archive at the Space Telescope Science Institute (STScI). Funding for the TESS mission is provided by the NASA Explorer Program. STScI is operated by the Association of Universities for Research in Astronomy, Inc., under NASA contract NAS 5–26555. LN and LM acknowledge the financial contribution from the INAF Large Grant 2023 ``EXODEMO''.
\end{acknowledgements}

%
\bibliographystyle{aa} 
\bibliography{references} 

\begin{appendix}
\nolinenumbers

\section{In-transit radial velocities}
\label{sec:appA}
The in-transit spectroscopic time series are available in electronic format as supplementary material at the CDS.

\section{Joint times series Bayesian analysis}
\label{sec:appB}
We examined our \textit{TESS}-corrected light curves and all spectroscopic time series within a Bayesian framework using \texttt{PyORBIT}\footnote{\url{https://github.com/LucaMalavolta/PyORBIT}} \citep{2016A&A...588A.118M, 2018AJ....155..107M}, a publicly accessible software that allows the modelling of planetary transits and RVs while considering the effects of stellar activity.

\begin{figure}[h]
   \centering
   \includegraphics[width=0.85\hsize]%
   {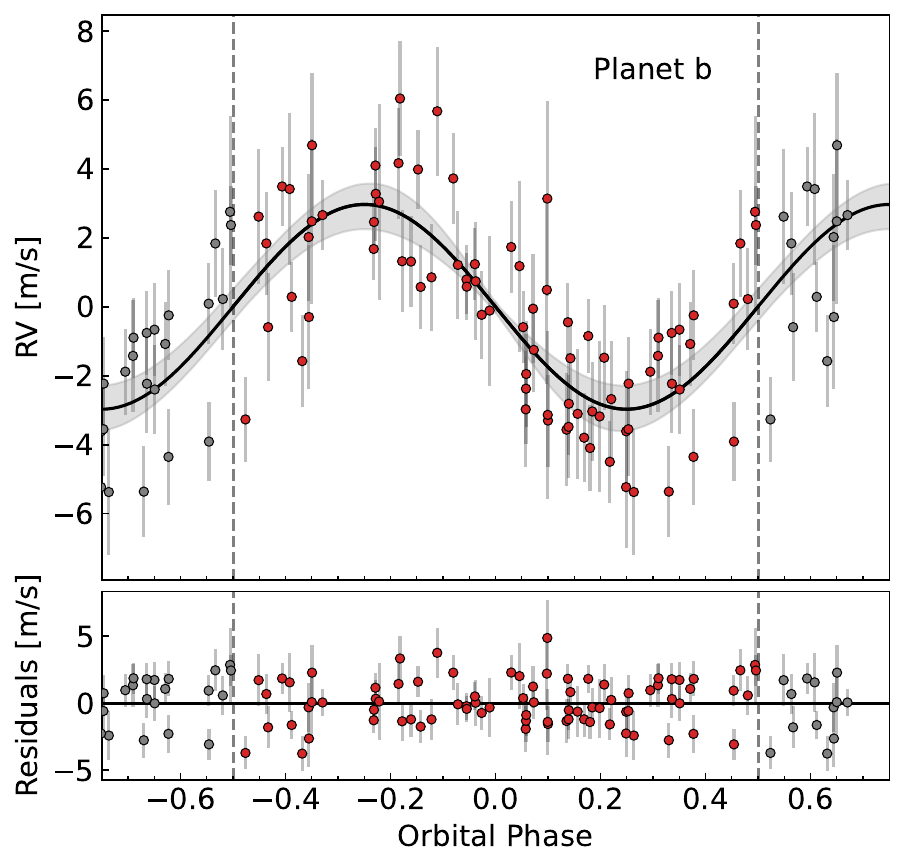}
   \caption{Phase-folded RV curve of TOI-1710 b. The grey-shaded area shows the $\pm 1\sigma$ uncertainties of the model, while the residuals are shown in the bottom panel. The fitted stellar activity and polynomial function are subtracted from the datapoints. }
   \label{fig:RV}
\end{figure}

First of all, we carefully considered the influence of stellar contamination from neighbouring stars, and verified stellar dilution by computing a dilution factor -- the total flux from contaminants falling into the photometric aperture divided by the flux contribution from the target star. We performed the calculation following \citet{2022MNRAS.516.4432M} and imposed a Gaussian prior in the modelling described as follows.  

In particular, we simultaneously modelled all \textit{TESS}-corrected light curves, the off-transit RV, BIS, and FWHM series, and the in-transit RV anomaly. This allowed us to model the planetary transits, the Keplerian signal and orbital obliquity, and the stellar activity using Gaussian processes \citep[GPs, ][and references therein]{2015MNRAS.452.2269R}. We used \texttt{BATMAN} \citep[][]{2015PASP..127.1161K} to model the planetary transits. We fitted the following parameters: central time of transit ($T_{0,~b}$), orbital period ($P_b$), planet to star radius ratio ($R_{\rm p}/R_\star$), impact parameter ($b$), RV semi-amplitude ($K_b$), and sky-projected obliquity ($\lambda$). Additionally, we calculated the eccentricity ($e$) and argument of pericenter ($\omega$), by fitting $\sqrt{e}\cos\omega$ and $\sqrt{e}\sin\omega$ \citep{2013PASP..125...83E}. We imposed Gaussian priors on the host star density ($\rho_\star$) and the projected rotational velocity ($v \sin {i_\star}$), while leaving the stellar rotation period ($P_{\rm rot}$, see below) free to vary. In particular, we treated the stellar equatorial velocity ($v_{\rm eq}$, obtained from the stellar radius $R_\star$ and $P_{\rm rot}$) and inclination ($i_\star$) as independent variables. Thus, we are not subject to the bias described in \citet{2020AJ....159...81M}. The $v \sin {i_\star}$ is a derived parameter; specifically, it is calculated from $v_{\rm eq}$ and $i_\star$ parameters at each step, and then compared with the given prior. We treated the stellar limb darkening (LD) contribution by estimating $\rm u_1$ and $\rm u_2$ using \texttt{PyLDTk}\footnote{\url{https://github.com/hpparvi/ldtk}} \citep{2013A&A...553A...6H, 2015MNRAS.453.3821P} and assuming a boxcar filter as the passband in the HARPS-N spectral range and adding $0.1$ in quadrature to their Gaussian errors to account for unknown model systematics. We also modelled and fit the effect of stellar convective blueshift (CB) on the in-transit RV curve \citep[e.g.,][]{2016A&A...588A.127C} using a linear law as a function of limb angle. Short-term stellar activity was included in the model by a jitter term added in quadrature to the RV and photometry errors.

Priors on stellar parameters (including limb-darkening coefficients) are based on posterior distributions from independent analyses, i.e., using data not involved in the fit, as described in Appendix \ref{sec:rot} and Table \ref{table:model-rm}. Parameters that cannot be measured directly from the RM effect, such as the stellar radius or the stellar inclination, are included in the fit so that they can be combined with other parameters constrained by the radial velocity and photometry (such as the stellar rotation period) and produce self-consistent derived parameters, such as the projected equatorial velocity, for the RM modelling. We would like to emphasise that we checked the robustness of our results by conducting a second analysis without imposing any prior information on vsini (leaving it free to vary). We obtained an almost identical posterior distribution, meaning that the chosen prior was not strongly informative. For all parameters, we employed uninformative boundaries based on physically realistic limits for a star of this spectral type. Specifically for the stellar rotation period, we kept this range wide to explore the parameter space, specifically to test the possible 11-day period solution. For some parameters, we restricted the parameter space while keeping it uninformative to reduce the convergence time. When boundaries are not specified in the table, we kept the default range (conservatively extremely wide) provided by \texttt{PyORBIT}.

We modelled stellar activity in the RV, BIS, and FWHM series using a multidimensional GP, while we used a unidimensional one for the \textit{TESS} photometric time series. We used an exponential-sine periodic (ESP) kernel as defined in \citet{2020A&A...638A..95D, 2022A&A...659A.182D} for the multidimensional GP, and a rotation kernel as defined in \citet{celerite1, celerite2}. As part of this modelling, we set the rotation period ($P_{\rm rot}$) as a unique parameter that is shared between the unidimensional GP, the multidimensional GP and the RM modelling. The rest of the GP hyperparameters, such as the characteristic decay timescale ($P_{\rm dec}$) and the coherence scale ($\omega_{\rm cs}$), are independent between the photometric and spectroscopic datasets. Finally, we included a 3rd degree polynomial function to model a long-term sinusoidal-like trend present in the spectroscopic time series. This function has coefficients shared among all datasets and a multiplication factor associated with each time series. 

We performed a global optimisation of the parameters by running \texttt{PyDE} \citep{Storn1997, 2016zndo.....45602P} for 100~000 generations and a Bayesian analysis of the RM signal in the RV time series using \textsc{emcee} \cite{2013PASP..125..306F} for 200~000 steps. This choice was motivated by the fact that MCMC samplers allow the specification of a prior on derived parameters, as in the case of $v \sin{i_\star}$, which is not possible in the case of nested sampling algorithms. We used $4\times n_{\rm dim}$ walkers, with $n_{\rm dim}$ the model dimensionality, and discarded the first 50~000 steps (burn-in). We applied a thinning factor of 100 to mitigate the effect of chain autocorrelation.  

\begin{figure}
   \centering
   \includegraphics[trim={0.2cm 0.2cm 0.2cm 0.2cm},clip,width=0.9\hsize]%
   {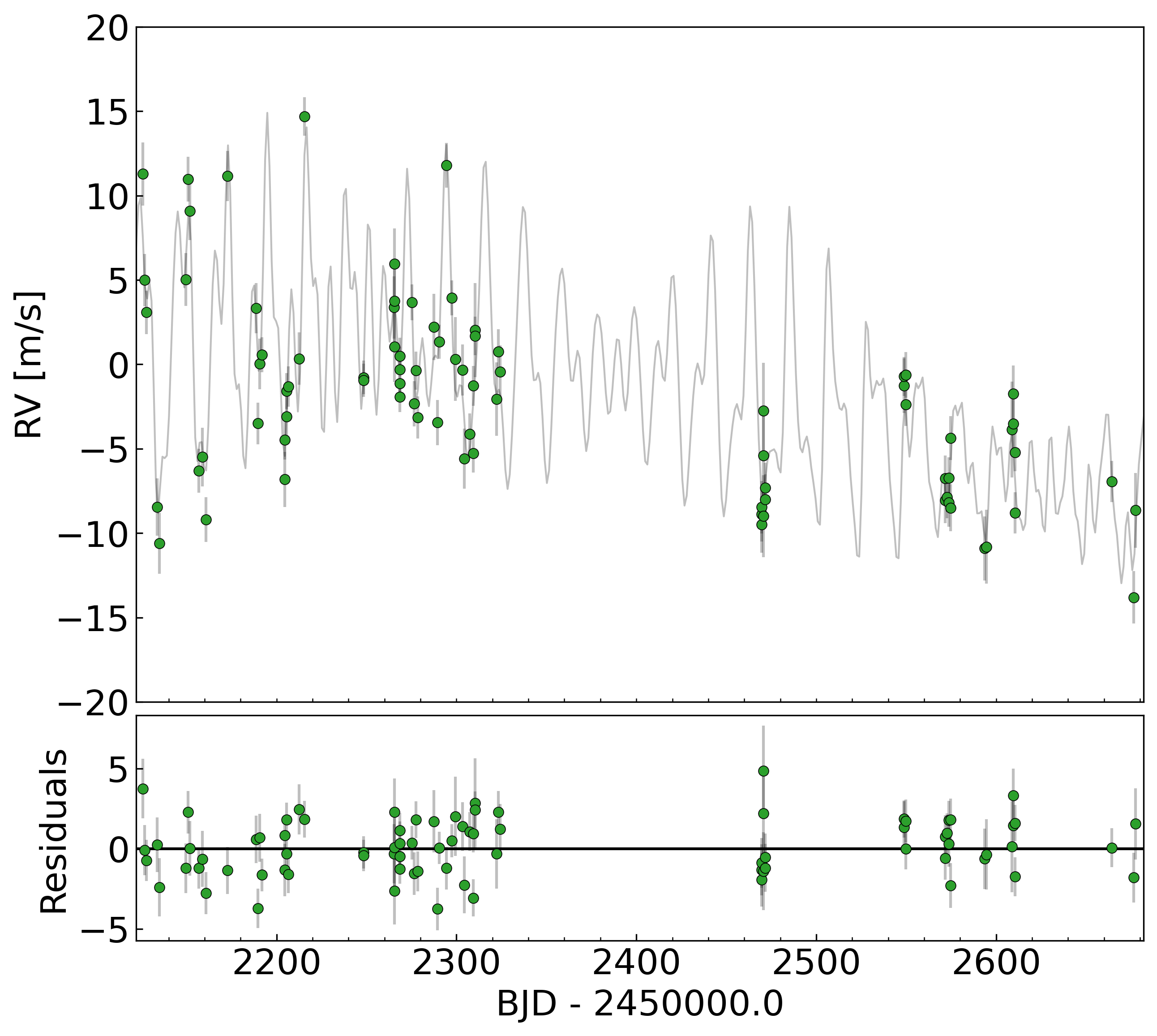}
   \caption{Modelling of HARPS-N RV time series. \textit{Top:} RV time series with superimposed full Keplerian $+$ stellar activity $+$ 3rd polynomial function model. \textit{Bottom:} Residuals of the fit.}
   \label{fig:rv_full}
\end{figure}

\section{Stellar rotation period refinement}
\label{sec:rot}

\begin{figure}[h]
   \centering
   \includegraphics[width=\hsize]%
   {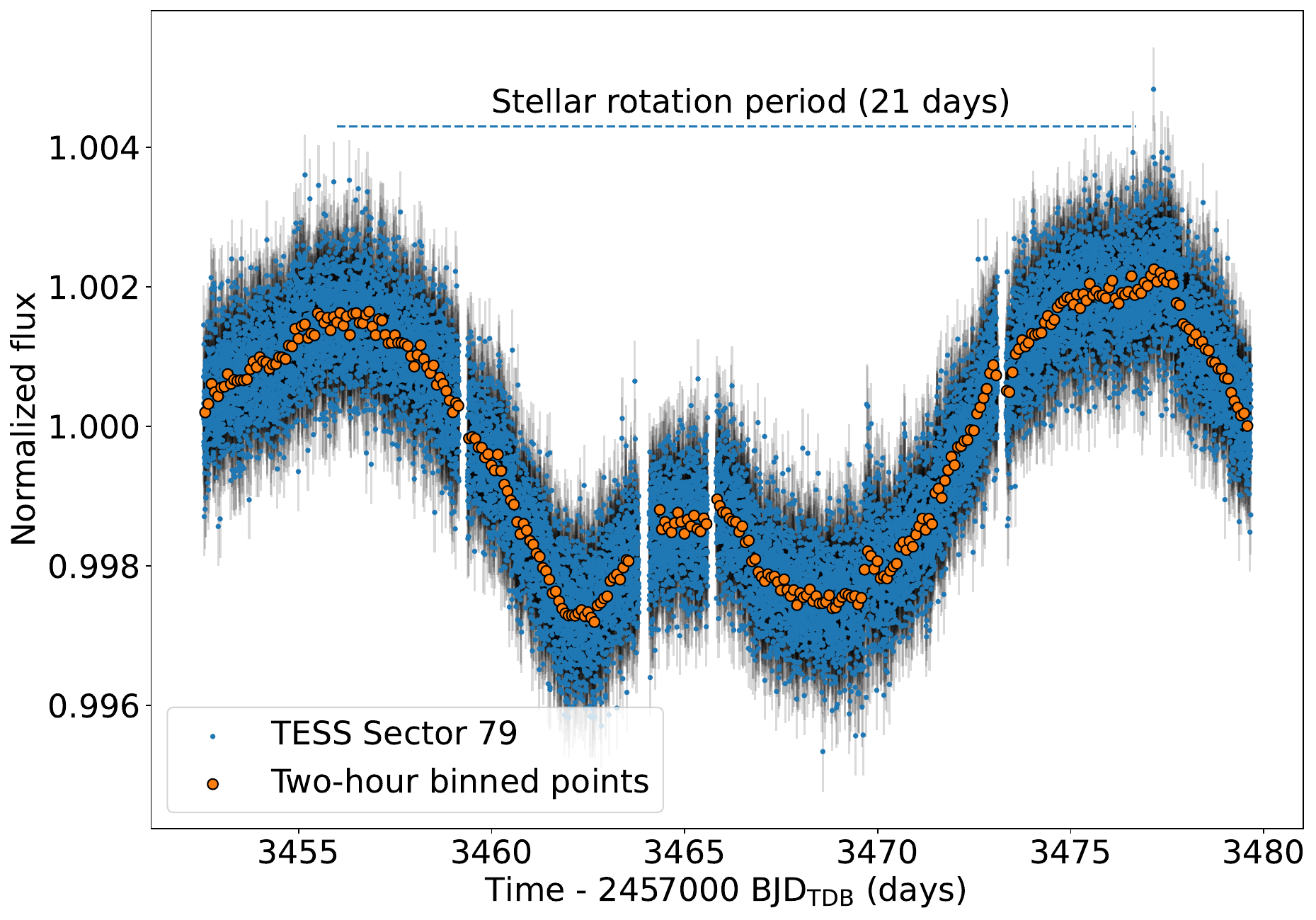}
   \caption{Photometric time series of \textit{TESS} sector 79. The 21-day stellar modulation can undoubtedly be seen. }
   \label{fig:s79}
\end{figure}

To accurately translate the sky-projected obliquity ($\lambda$) into the true 3D orbital obliquity ($\psi$), we need a reliable stellar rotation period. We estimated this in our fully Bayesian analysis by leaving it free to vary and modelling the activity in our \textit{TESS}-corrected light curves (which preserve stellar variability), RV, BIS, and FWHM series. This yielded a stellar rotation period $P_{\rm rot}\, = 21.5 \pm 0.2$~days, which is compatible with that found in \citetalias{2022AA...666A.183K} and \citet{2024AA...684A..96O}.  
To support our finding, we used the Generalised Lomb-Scargle (GLS) algorithm \citep{2009A&A...496..577Z} to examine the periodogram of the two spectroscopic activity indicators we used (BIS, FWHM) and the RVs. The periodogram of the BIS shows a dominant peak close to 11 days, while the other two close to 21 days. Most \textit{TESS} sectors also show the highest peak around 10-11 days, with occasional signals close to 17 days. However, sector 79 agrees with previous works, showing the dominant peak at 21 days (Fig. \ref{fig:s79}). This stellar variability cannot be physically explained by a faster rotation with a periodicity of half that of the 21-day. Instead, it is possible to observe a shorter period in other lightcurves if there are two similarly sized spots in opposite hemispheres.

Given these considerations, we adopt as the most reliable $P_{\rm rot}$ that obtained from our full analysis (see Table \ref{table:model-rm}), interpreting the 11-day signal as the first harmonic of the stellar rotation. This interpretation is further supported by multiple diagnostics, including the stellar kinematic properties and lithium content, as detailed in the following section.

\section{Independent stellar age diagnostics}
\label{app_age}

First of all, a rotational period of 11 days would imply a gyrochronological age of about 900 Myr, according to the calibration in \citet{mamajek2008}. However, the kinematic properties of the star disfavour such a moderately young age. In fact, the space velocities, derived following \citet{johnson1987}, are U = $44.79 \pm 0.14$ km\,s$^{-1}$, V = $-20.17\pm0.12$ km\,s$^{-1}$, and W = $8.87\pm0.09$ km\,s$^{-1}$. These values are well outside the kinematic space in which nearby young ($\leq$ 1 Gyr) stars are confined \citep{montes2001}. The probability distribution function of the age, evaluated on the basis of the stellar kinematics according to the method in \citet{2018MNRAS.476..184A}, has a maximum at about 4.6 Gyr and gives an estimated kinematic age of $5.2 \pm 3$~Gyr, thus supporting the results based on \citet{montes2001}. A similar discrepancy would arise for the mean chromospheric activity level $\log R'_{HK} \approx -4.78$ \citep{2022AA...666A.183K, 2024AA...684A..96O}. This differs greatly from what would be expected for a period of 11 days and an age of about 900 Myr (about $-4.4$ for such G dwarf stars). Conversely, the gyrochronological age resulting from the adopted rotation period, i.e. about 3 Gyr, is instead fully consistent with the age derived from the chromospheric activity and stellar kinematics.

The co-added HARPS-N spectrum of the target show the lithium line at $\sim$6707.8\,\AA\, with a mean equivalent width $EW_{\rm Li}=5.7\pm0.5$ m\AA. This corresponds to a lithium abundance $\log n({\rm Li}) = 1.20\pm0.05$ dex, corrected for non-LTE effects \citep{2009A&A...503..541L} following the spectral synthesis method detailed in \citet{2022A&A...664A.161B}. Based on the effective temperature reported in \citetalias{2022AA...666A.183K}, the methodology described in \citet{jeffries_2023} and the derived $EW_{\rm Li}$, we estimate a lower age limit of $ > 1.3$\,Gyr for the target. Conversely, its lithium abundance, when placed in a $\log n({\rm Li})$-$T_{\rm eff}$ diagram, is compatible with the lower envelope of M67 members, which implies an upper age limit of $<4.5$\,Gyr \citep{SestitoRandich2005}. This result further supports the 21.5-day rotation period, as a half-periodicity would imply an age younger than the lower limit of 1.3 Gyr. It is worth noting that our results agree well with previous age estimates for this system, consistently indicating an age greater than 1.3 Gyr.

To support our findings, we have also computed $P_{\rm rot}$ adopting the equations reported in \cite{1984ApJ...287..769N}. Assuming $\log R'_{HK} = -4.78 \pm 0.03$ \citepalias{2022AA...666A.183K} and $B-V = 0.66 \pm 0.03$ \citep{2000A&A...355L..27H}, we found the theoretically-derived convective overturn time to be $\tau_c = 12.6 \pm 1.7$~d, and $P_{\rm rot} = 20.3 \pm 3.0$~d, in agreement with the value reported in Table \ref{table:model-rm}.

\section{The inflated nature of TOI-1710~b}
The inflated nature of TOI-1710~b, coupled with its low incident flux and late stellar age (see sect. \ref{app_age}), suggests it may have resisted standard interior cooling. We propose that tidal heating, produced during the final stages of HEM, may have served as a later energy source \citep[see also][]{2017MNRAS.464.3937G}. In line with thermal evolution models by \citet{2014ApJ...792....1L} and previous studies on similarly inflated planets orbiting old stars \citep{2011ApJS..197....7C}, the inflation of TOI-1710~b may also be attributed to a low mass fraction of heavy elements and a large mass fraction of H-He gas. This would make it more likely to host an extended atmosphere despite its proximity to the star.

The HEM process requires a remarkable decrease of the mechanical energy of the initial planetary orbit that is produced by tidal dissipation inside the planet \citep[e.g.][]{2018ARAA..56..175D}. Specifically, the energy decrease, starting from an initial orbit with a semimajor axis much larger than the present value,  is given by $E_{\rm diss} = (1/2) GM_{\rm s} M_{\rm p}/a$, where $G$ is the gravitation constant, $M_{\rm s}$ the mass of the star, $M_{\rm p}$ the mass of the planet, and $a$ the present orbit semimajor axis. The system parameters give $E_{\rm diss} \sim 2 \times 10^{35}$~J. The total duration of the HEM process and, therefore, of the dissipation of such a large amount of energy, can be comparable with the age of the system. Assuming, for example, a duration of 1.0 Gyr, we get an average dissipated power of $\sim 7 \times 10^{18}$~W. Even assuming that all that power is used to evaporate an H/He envelope having the mass of 1\% of that of the planet, the lifetime of the envelope is of the order of $\sim 1.5$~Gyr, that is, comparable with the age of the star. Nevertheless, a remarkable fraction of the tidal energy should go into radiation from the planetary atmosphere, thus substantially increasing the evaporation timescale of the envelope. On the other hand, such an internal tidal heating, extended over a timescale comparable with the age of the system, can effectively oppose the contraction of the planet associated with its cooling after formation \citep[e.g. ][]{2014ApJ...792....1L}, thus keeping its structure inflated. 

At the same time, the HEM scenario that we propose offers a plausible explanation for TOI-1710~b to have escaped atmospheric erosion. Planets undergoing disc-migration typically arrive in close-in orbits within 10 Myr, exposing them to the full, constant XUV irradiation of the star in its active phase. In contrast, planets undergoing late HEM may escape this phase \citep[e.g. ][]{2018Natur.553..477B, 2021AA...647A..40A}. Although the XUV flux at pericenter is significantly higher than during disc-migration (potentially four times that of the final circular orbit), the planet spends most of its time at apocenter, safe from intense XUV flux irradiation. A planet on a highly eccentric orbit ($e \approx$ 0.8) spends only about 1.2\% of its orbital period within $\pm$ 30 deg of pericenter (5\% when $e \approx$ 0.5). This reduced exposure likely enabled TOI-1710~b to survive as a gas-rich world. Another possible explanation for the photo-evaporation resistance and inflation of TOI-1710 b is in the metal-rich nature of TOI-1710 (see Tab. \ref{table:star}). This possible correlation was first noted by \citet{2022MNRAS.511.1043W}, which postulated that planets orbiting metal-rich stars have metal-rich atmospheres with reduced photo-evaporation.

\section{Search for planetary companions}
\label{app_twothree}
We tested for the presence of additional planetary companions by performing the same Bayesian analysis detailed in Sect. \ref{sec:appB}, including one or two additional Keplerian signals. Specifically, we imposed wide uniform priors on their orbital periods (0 -- 2000 days), eccentricities and RV semi-amplitudes (0.01 -- 50 m\,s$^{-1}$). This further test was performed due to the large obliquity found and its dynamical implications, which are discussed in the following section. Moreover, the analysis was motivated by the relatively large value of the uncorrelated RV jitter (1.4 m\,s$^{-1}$).  

The two-planet solution gives a similar jitter value and finds a candidate signal at approximately 1650 days with a high eccentricity ($>$ 0.8). Instead, the three-planet model results in a slightly lower jitter of 1.3 m\,s$^{-1}$ and suggests two different candidate signals with periods of about 1480 and 105 days, both detected at 3$\sigma$. The period of the wide orbit candidate in this model is inconsistent with the one found in the two-planet solution, which suggests that the corresponding RV signal is more likely to be of stellar origin, and that it could not be perfectly modelled with the polynomial trend. We repeated these analyses with the 3rd degree polynomial function removed. As before, the long-period candidate changes its periodicity in the two- and three-planet solutions. However, the period and amplitude of the third signal remained compatible.

To compare the one-, two- and three-planet solutions, we used the Bayesian information criterion \citep[BIC,][]{1978AnSta...6..461S}. We performed the comparison separately for the models with and without the polynomial function included. As shown in Table \ref{table:bic_comparison}, the one-planet solution is strongly preferred in both cases. It is worth noting that the Keplerian and obliquity solutions for planet b show little variation (less than one sigma) across the tested models, further strengthening the results (see Appendix. \ref{sec:appB}) and conclusions of this work.

\begin{table}
\caption{BIC values for the different models tested.}      
\label{table:bic_comparison}   
\centering                      
\begin{tabular}{c c c c}      
\hline\hline                 
Model & Polynomial trend & BIC & $N_{\rm params}$ \rule{0pt}{2.5ex} \rule[-1ex]{0pt}{0pt} \\    
\hline
1--planet & yes & $-$\textbf{112086.1} & 58\rule{0pt}{2.5ex} \rule[-1ex]{0pt}{0pt}\\ 
2--planet & yes & $-$112060.6 & 63 \rule{0pt}{2.5ex} \rule[-1ex]{0pt}{0pt}\\ 
3--planet & yes & $-$112037.5 & 68 \rule{0pt}{2.5ex} \rule[-1ex]{0pt}{0pt}\\ 
\hline                       
1--planet & no  & $-$\textbf{112115.9} & 53\rule{0pt}{2.5ex} \rule[-1ex]{0pt}{0pt}\\ 
2--planet & no  & $-$112089.8 & 58 \rule{0pt}{2.5ex} \rule[-1ex]{0pt}{0pt}\\ 
3--planet & no  & $-$112070.0 & 63 \rule{0pt}{2.5ex} \rule[-1ex]{0pt}{0pt}\\ 
\hline                                 
\end{tabular}
\end{table}

\section{Detection completeness map}
\label{sect:det}
To quantify our sensitivity to additional companions, we performed an injection-recovery analysis, based on the HARPS-N RVs, following the same approach used in \citet{2025AA...701A..79N}. We explored a logarithmic grid in planetary minimum mass, ranging from $10^{-1}\,M_{\rm Jup}$ to $20\,M_{\rm Jup}$, and semi-major axis ($a$), from $10^{-2}$ to $10$ au. For each grid cell, we generated 200 synthetic planetary systems. The orbital period was derived from $a$, while the time of conjunction and argument of periastron were drawn from uniform distributions. The inclination was sampled from a sinusoidal distribution (i.e. uniform in $\cos i$), and the eccentricity was drawn from a Beta distribution following \citet{2013MNRAS.434L..51K}, representative of the observed exoplanet population.

\begin{figure}
   \centering
   \includegraphics[width=0.9\hsize]%
   {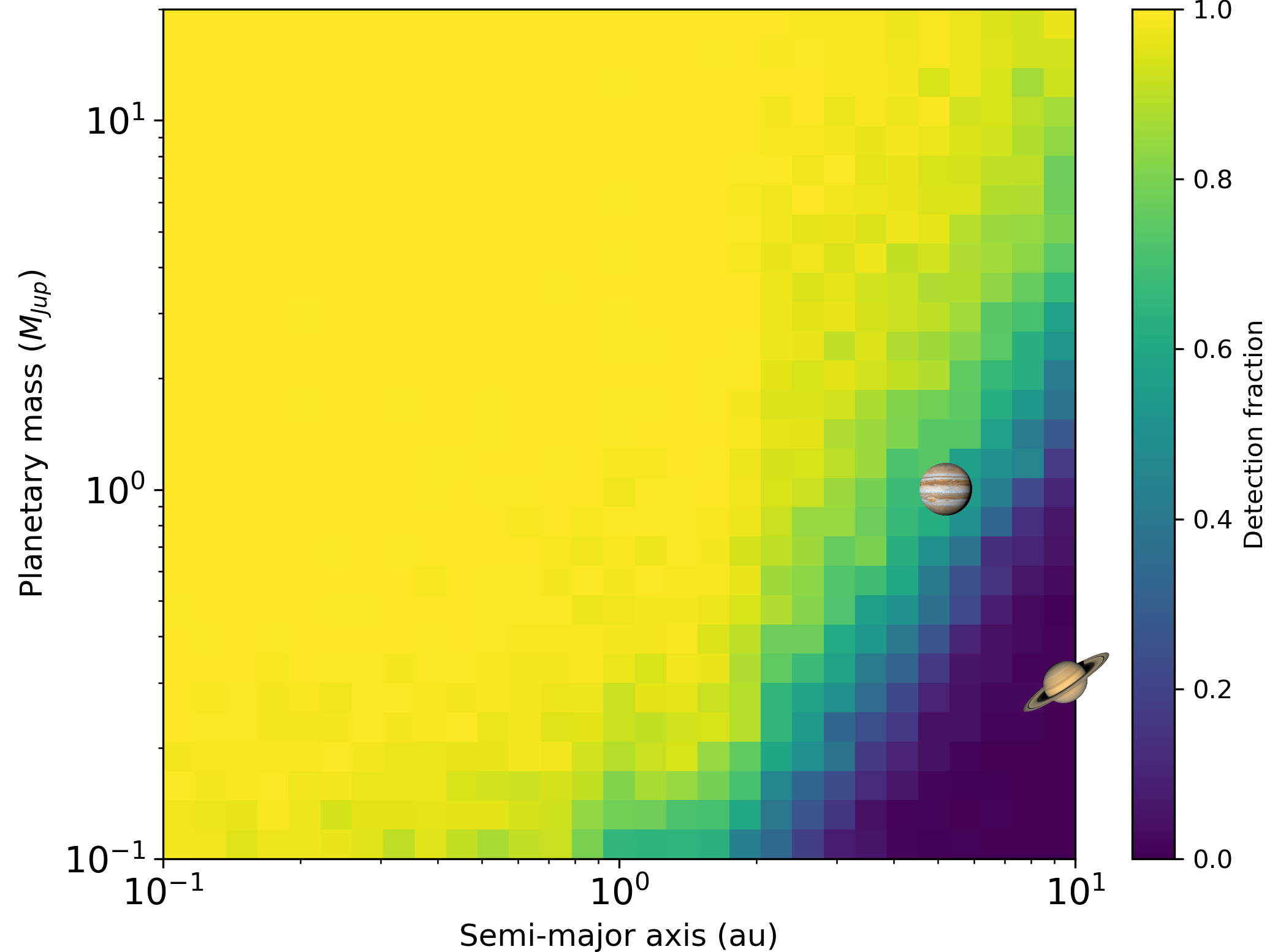}
   \caption{Detection completeness map for TOI-1710 in the mass versus semi-major axis plane, based on the RV injection-recovery analysis. The colour scale represents the fraction of simulated planetary companions recovered in each grid cell. For reference, the locations of Jupiter and Saturn are shown as icons.}
   \label{fig:completeness}
\end{figure}

For each simulated planet, we generated synthetic RV time series at the actual observing epochs and added Gaussian noise scaled to the instrumental jitter reported in Table\,\ref{table:model-rm2}. We then compared three models: (i) a constant model, (ii) a polynomial trend (linear or quadratic) to account for long-period variability, and (iii) a full Keplerian model. Model selection was performed using the Bayesian Information Criterion (BIC), defined as $\mathrm{BIC} = \chi^2 + k\ln N_{\rm obs}$, where $k$ is the number of free parameters and $N_{\rm obs}$ the number of measurements. A planet was considered detected if the Keplerian model was preferred over the simpler models with $\Delta \mathrm{BIC} > 10$ \citep{Kass1995}. In addition, signals producing significant linear or quadratic trends were flagged as detected when the polynomial model was favoured over the constant one. The completeness map (Fig.\,\ref{fig:completeness}) reports, for each grid cell, the fraction of recovered planets, defined as $N_{\rm det}/N_{\rm sim}$. With the current HARPS-N dataset, our sensitivity allows us to robustly detect Jupiter-mass companions out to approximately $\sim4$ au, while lower-mass planets remain detectable only at shorter orbital separations.

\section{Probing photo-evaporation via HeI triplet and H$\alpha$}
\label{app:photo}
We conducted a multi-band transmission spectroscopy analysis to better understand the atmospheric properties of TOI-1710 b. In particular, we investigated two key tracers of extended and possibly escaping atmospheres: the HeI triplet at 1083.3\,nm (GIANO-B data) and the H$\alpha$ line \citep[HARPS-N data,][]{darpa_toi5398}. We carried out HeI and H$\alpha$ transmission spectroscopy by comparing in- and out-of-transit spectra, following the methodology described in \citet{Guilluy2024} and \citet{darpa_toi5398}. 

We did not detect any clear absorption features at the positions of the HeI triplet or the H$\alpha$ line. Following the method of \citet{Guilluy2024}, we report 1$\sigma$ upper limits on the excess absorption at the expected line positions of 0.58$\%$ for HeI and 0.29$\%$ for H$\alpha$, corresponding to effective planetary radii of 1.8~$R_\mathrm{b}$ and 1.5~$R_\mathrm{b}$, respectively. The non-detection of HeI and H$\alpha$ for TOI-1710\,b is not surprising. The planet's relatively low equilibrium temperature ($T_{\rm eq} \approx 670$~K), combined with its long orbital period, may reduce the effects of stellar irradiation, inhibiting the population of the $n = 2$ level from which the H$\alpha$ transition originates. The same applies to HeI, where the planetary distance from the star may reduce the XUV flux, disfavouring the mechanisms of ionisation and recombination that leads to formation of the metastable HeI triplet.

\section{Extra figures and tables}
\label{sec:appC}

\begin{figure}
   \centering
   \includegraphics[trim={1.7cm 0.9cm 1.4cm 1.8cm},clip,width=0.8\hsize]%
   {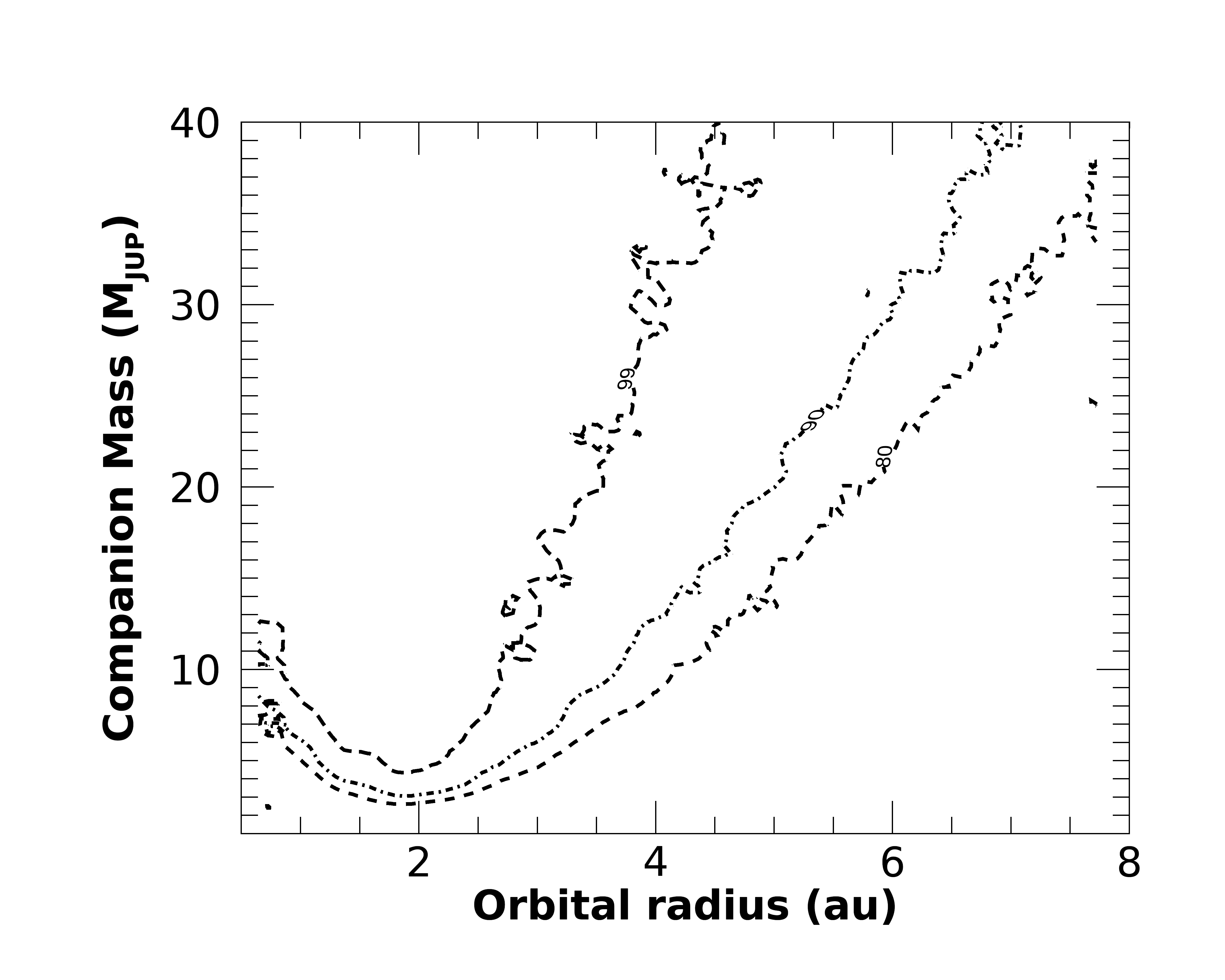}
   \caption{Astrometric Gaia DR3 sensitivity map for companions. The dashed, dotted and long-dashed lines correspond to isoprobability curves for 80, 90 and 99\% probability of a companion with given properties producing RUWE $>$ 0.94.}
   \label{fig:astrometry}
\end{figure}

\begin{figure}
   \centering
   \includegraphics[width=0.8\linewidth]%
   {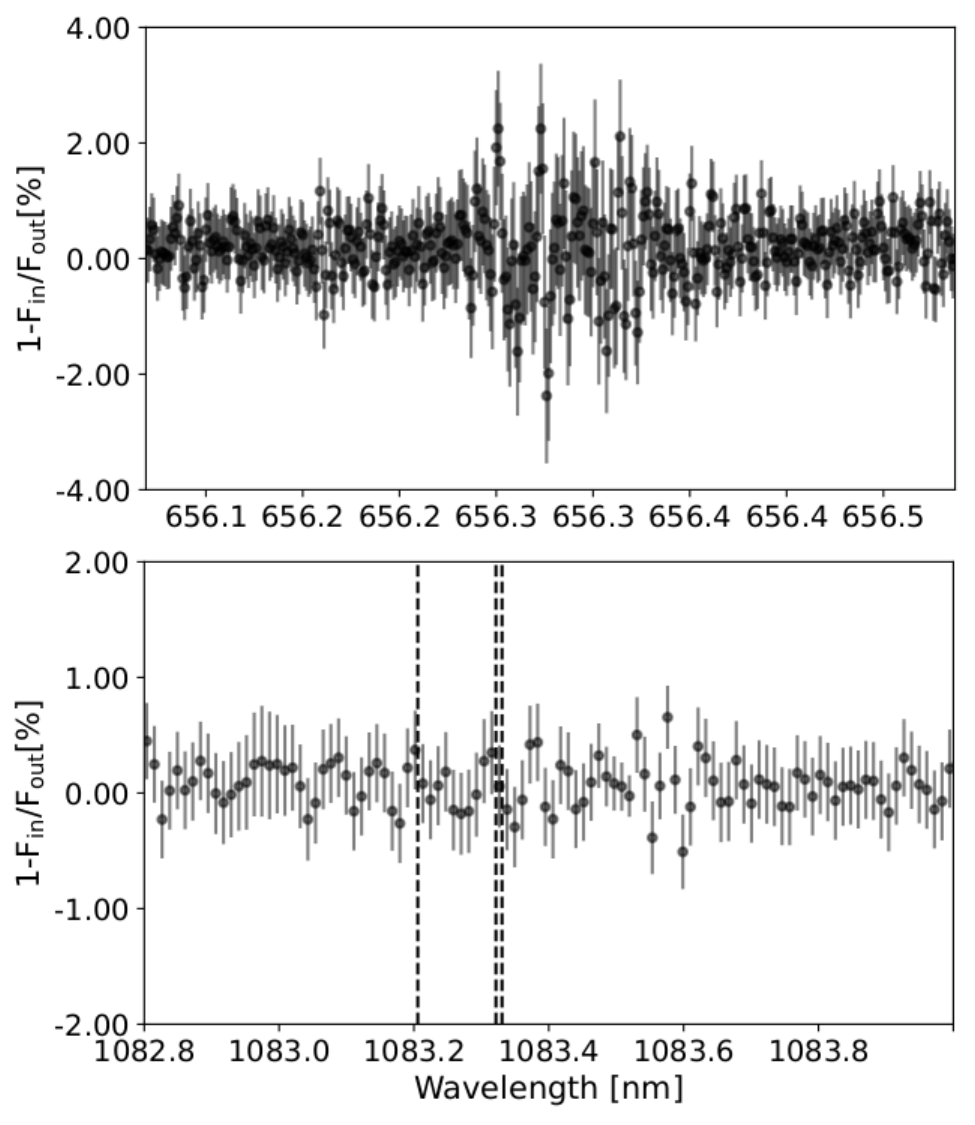}
   \caption{Final transmission spectrum around the H$\alpha$ (top) and HeI (bottom) evaporation diagnostics. In the right panel, vertical black dotted lines indicate the position of the HeI triplet.}
   \label{fig:he}
\end{figure}

\begin{table}
\caption{Priors and outcomes of RM modelling.}             
\label{table:model-rm}
\addtolength{\tabcolsep}{-0.75em}
\centering          
\begin{tabular}{l c c c} 
\hline\hline     
 
Parameter & Unit & Prior & Value \rule{0pt}{2.3ex} \rule[-1ex]{0pt}{0pt}\\ 
\hline 
\multicolumn{4}{c}{Planetary parameters} \rule{0pt}{2.3ex} \rule[-1ex]{0pt}{0pt}\\ 
\hline
   Orbital period ($P_{\rm b}$) & days & $\mathcal{U}$(24.282,  & 24.28336(1) \rule{0pt}{2.3ex} \rule[-1ex]{0pt}{0pt}\\
    &  & 24.284) &  \rule{0pt}{2.3ex} \rule[-1ex]{0pt}{0pt}\\
   Central transit time ($T_{\rm 0}$) & BTJD\tablefootmark{a} & $\mathcal{U}$(2031.1,  & 2031.2301(4) \rule{0pt}{2.3ex} \rule[-1ex]{0pt}{0pt}\\
    &  & 2031.3) &  \rule{0pt}{2.3ex} \rule[-1ex]{0pt}{0pt}\\
   Scaled semi-maj. axis ($\frac{a_{\rm b}}{R_{\star}}$) &  & ... & 36.3$^{+0.6}_{-0.7}$ \rule{0pt}{2.3ex} \rule[-1ex]{0pt}{0pt}\\
   Orbital semi-maj. axis ($a_{\rm b}$) & au & ... & 0.163$\pm$0.003 \rule{0pt}{2.3ex} \rule[-1ex]{0pt}{0pt}\\
   Orbital inclination ($i_{\rm b}$) & deg & ... & 89.82$^{+0.12}_{-0.17}$ \rule{0pt}{2.3ex} \rule[-1ex]{0pt}{0pt}\\
   Orbital eccentricity ($e_{\rm b}$) &  & $\mathcal{U}$(0.0, 0.95) & 0.04$^{+0.07}_{-0.03}$ 
   \rule{0pt}{2.3ex} \rule[-1ex]{0pt}{0pt}\\
   RV semi-amplitude ($K_{\rm b}$) & m s$^{-1}$ & $\mathcal{U}$(0.001, 100) & 2.97$^{+0.62}_{-0.64}$ \rule{0pt}{2.3ex} \rule[-1ex]{0pt}{0pt}\\
   Planet/star rad. ratio ($\frac{R_{\rm p}}{R_{\star}}$) &  & $\mathcal{U}$(0.0, 0.5) & 0.04979(35) \rule{0pt}{2.3ex} \rule[-1ex]{0pt}{0pt}\\
   Impact parameter ($b$) &  & $\mathcal{U}$(0.0, 2.0) & 0.109$^{+0.11}_{-0.076}$ \rule{0pt}{2.3ex} \rule[-1ex]{0pt}{0pt}\\
   Transit duration ($T_{14}$) & days & ... & 0.2216$^{+0.0044}_{-0.0034}$ \rule{0pt}{2.3ex} \rule[-1ex]{0pt}{0pt}\\
   Planetary radius ($R_b$) & $R_\oplus$ & ... & 5.258$\pm$0.085 \rule{0pt}{2.3ex} \rule[-1ex]{0pt}{0pt}\\
   Planetary mass ($M_b$) & $M_\oplus$ & ... & 13.2$^{+2.8}_{-2.9}$ \rule{0pt}{2.3ex} \rule[-1ex]{0pt}{0pt}\\
   Planetary density ($\rho_b$) & g cm$^{-3}$  & ... & 0.50$\pm$0.11 \rule{0pt}{2.3ex} \rule[-1ex]{0pt}{0pt}\\
   Arg. of pericenter ($\omega$) & deg & ... & 126$^{+67}_{-119}$ \rule{0pt}{2.3ex} \rule[-1ex]{0pt}{0pt}\\
   Sky-proj. obliquity ($\lambda$) & deg & $\mathcal{U}$(0.0, 360) & 176$^{+29}_{-27}$ \rule{0pt}{2.3ex}  \rule[-1ex]{0pt}{0pt}\\
   True 3D obliquity ($\psi$) & deg & ... & 149$^{+11}_{-10}$ \rule{0pt}{2.3ex} \rule[-1ex]{0pt}{0pt}\\
\hline

\multicolumn{4}{c}{Stellar parameters} \rule{0pt}{2.3ex} \rule[-1ex]{0pt}{0pt}\\ 
\hline 
   Density ($\rho_{\star}$) & $\rho_{\odot}$ & $\mathcal{N}$(1.08, 0.08)\tablefootmark{b} & 1.09$^{+0.05}_{-0.06}$   \rule{0pt}{2.3ex} \rule[-1ex]{0pt}{0pt}\\
   Radius ($R_{\star}$) & $R_{\odot}$ & $\mathcal{N}$(0.968, \tablefootmark{b} & 0.968$\pm$0.014   \rule{0pt}{2.3ex} \rule[-1ex]{0pt}{0pt}\\
    &  & 0.016)\tablefootmark{b} &    \rule{0pt}{2.3ex} \rule[-1ex]{0pt}{0pt}\\
   Rotation period ($P_{\rm rot}$) & days & $\mathcal{U}$(8.0, 24) & 21.48$\pm$0.20 \rule{0pt}{2.3ex} \rule[-1ex]{0pt}{0pt}\\
   Decay Timescale ($P_{\rm dec}$) & days & $\mathcal{U}$(40, 1000) & 69$^{+22}_{-16}$ \rule{0pt}{2.3ex} \rule[-1ex]{0pt}{0pt}\\
   Coherence scale ($\omega$) &  & $\mathcal{N}$(0.35, 0.035)\tablefootmark{c} & 0.39$\pm$0.03 \rule{0pt}{2.3ex}   \rule[-1ex]{0pt}{0pt}\\
   Rotat. velocity ($v \sin{i_\star}$) & km s$^{-1}$ & $\mathcal{N}$(2.3, 0.4)\tablefootmark{c} &  2.17$^{+0.10}_{-0.17}$   \rule{0pt}{2.3ex} \rule[-1ex]{0pt}{0pt}\\
   Quadratic LD coeff. ($u_1$) &  & $\mathcal{N}$(0.62, 0.11) & 0.63$\pm$0.11 \rule{0pt}{2.3ex} \rule[-1ex]{0pt}{0pt}\\
   Quadratic LD coeff. ($u_2$) &  & $\mathcal{N}$(0.11, 0.12) & 0.13$\pm$0.12 \rule{0pt}{2.3ex} \rule[-1ex]{0pt}{0pt}\\
   Stellar inclination ($i_{\star}$) & deg & $\mathcal{U}$(0, 180) & 72$^{+12}_{-11}$\rule{0pt}{2.3ex} \rule[-1ex]{0pt}{0pt}\\
   Convective blueshift (c$_1$) & & $\mathcal{U}$($-$2, 0)& $-$0.69$^{+0.46}_{-0.59}$\rule{0pt}{2.3ex} \rule[-1ex]{0pt}{0pt}\\
   Equatorial velocity ($v_{\rm eq}$) & km s$^{-1}$ & $\mathcal{U}$(0, 70) & 2.28$\pm$0.04 \rule{0pt}{2.3ex} \rule[-1ex]{0pt}{0pt}\\
\hline   
\end{tabular}
\tablefoot{
\tablefoottext{a}{BTJD = BJD$_{\rm TDB}$ - 2457000.0.}
\tablefoottext{b}{\citetalias{2022AA...666A.183K}.}
\tablefoottext{c}{\citet{2022AA...664A.163N}.}
}
\end{table}

\begin{table}
\caption{Literature stellar parameters}             
\label{table:star}      
\addtolength{\tabcolsep}{-0.3em}
\centering     
\begin{tabular}{l c c c c} 
\hline\hline     
Reference & Mass & T$_{\rm eff}$ & [Fe/H]  \rule{0pt}{2.3ex} \rule[-1ex]{0pt}{0pt}\\ 
 & $M_{\odot}$ & K & dex  \rule{0pt}{2.3ex} \rule[-1ex]{0pt}{0pt}\\ 
\hline
\citetalias{2022AA...666A.183K} & 0.9840$^{+0.05}_{-0.06}$ & 5665 $\pm$ 55 &  0.10(7)  \rule{0pt}{2.3ex} \rule[-1ex]{0pt}{0pt}\\
\citet{2023AJ....166...33M} & 0.957$^{+0.016}_{-0.015}$ & 5684$^{+64}_{-68}$ &  0.15(5)  \rule{0pt}{2.3ex} \rule[-1ex]{0pt}{0pt}\\
\citetalias{2024AA...684A..96O} & 0.95 $\pm$ 0.02 & 5730 $\pm$ 30 & 0.12(6)   \rule{0pt}{2.3ex} \rule[-1ex]{0pt}{0pt}\\
 \hline
\end{tabular}
\end{table}

\begin{table}
\caption{Priors and outcomes of full modelling.}             
\label{table:model-rm2}
\addtolength{\tabcolsep}{-0.8em}
\centering          
\begin{tabular}{l c c c c} 
\hline\hline     
Parameter & Unit & Prior & Value \rule{0pt}{2.3ex} \rule[-1ex]{0pt}{0pt}\\ 
\hline 
   \textit{TESS} s19 jitter ($\sigma^{\rm s19}_{jitter}$) & ppt & ... & 0.05$\pm$0.01 \rule{0pt}{2.3ex} \rule[-1ex]{0pt}{0pt}\\
   \textit{TESS} s20 jitter ($\sigma^{\rm s20}_{jitter}$) & ppt & ... & 0.090$\pm$0.006 \rule{0pt}{2.3ex} \rule[-1ex]{0pt}{0pt}\\
   \textit{TESS} s26 jitter ($\sigma^{\rm s26}_{jitter}$) & ppt & ... & 0.056$\pm$0.010 \rule{0pt}{2.3ex} \rule[-1ex]{0pt}{0pt}\\
   \textit{TESS} s40 jitter ($\sigma^{\rm s40}_{jitter}$) & $e^{-}$s$^{-1}$ & ... & 14.4$^{+2.2}_{-2.2}$ \rule{0pt}{2.3ex} \rule[-1ex]{0pt}{0pt}\\
   \textit{TESS} s53 jitter ($\sigma^{\rm s53}_{jitter}$) & ppt & ... & 0.055$\pm$0.010 \rule{0pt}{2.3ex} \rule[-1ex]{0pt}{0pt}\\
   \textit{TESS} s60 jitter ($\sigma^{\rm s60}_{jitter}$) & ppt & ... & 0.067$\pm$0.006 \rule{0pt}{2.3ex} \rule[-1ex]{0pt}{0pt}\\
   \textit{TESS} s73 jitter ($\sigma^{\rm s73}_{jitter}$) & ppt & ... & 0.056$\pm$0.011 \rule{0pt}{2.3ex} \rule[-1ex]{0pt}{0pt}\\
   \textit{TESS} s79 jitter ($\sigma^{\rm s79}_{jitter}$) & ppt & ... & 0.054$\pm$0.009 \rule{0pt}{2.3ex} \rule[-1ex]{0pt}{0pt}\\
Uncorr. RV jitter ($\sigma^{\rm RV}_{jitter}$) & m s$^{-1}$ & ... & 1.4$^{+0.4}_{-0.3}$   \rule{0pt}{2.3ex} \rule[-1ex]{0pt}{0pt}\\
Uncorr. RM jitter ($\sigma^{\rm RM}_{jitter}$) & m s$^{-1}$ & ... & 0.4$^{+0.4}_{-0.3}$   \rule{0pt}{2.3ex} \rule[-1ex]{0pt}{0pt}\\
Uncorr. BIS jitter ($\sigma^{\rm BIS}_{jitter}$) & m s$^{-1}$ & ... & 2.7$\pm$0.6   \rule{0pt}{2.3ex} \rule[-1ex]{0pt}{0pt}\\
Uncorr. FWHM jitter & km s$^{-1}$ & ... & 0.0045$\pm$0.0007   \rule{0pt}{2.3ex} \rule[-1ex]{0pt}{0pt}\\
RV offset ($\gamma^{\rm RV}$) & m s$^{-1}$ & ... & $-$38814.5$^{+1.9}_{-2.5}$  \rule{0pt}{2.3ex} \rule[-1ex]{0pt}{0pt}\\
RM offset ($\gamma^{\rm RM}$) & m s$^{-1}$ & ... & $-$38819.5$\pm$0.4   \rule{0pt}{2.3ex} \rule[-1ex]{0pt}{0pt}\\
BIS offset ($\gamma^{\rm BIS}$) & m s$^{-1}$  & ... & $-$62.6$^{+1.7}_{-2.0}$   \rule{0pt}{2.3ex} \rule[-1ex]{0pt}{0pt}\\
FWHM offset ($\gamma^{\rm FWHM}$) & km s$^{-1}$ & ... & 7.257$^{+0.006}_{-0.007}$   \rule{0pt}{2.3ex} \rule[-1ex]{0pt}{0pt}\\
\textit{TESS} quad. LD coeff. ($u_1$) &  & $\mathcal{N}$(0.42, 0.1) & 0.45$\pm$0.05 \rule{0pt}{2.3ex} \rule[-1ex]{0pt}{0pt}\\
\textit{TESS} quad. LD coeff. ($u_2$) &  & $\mathcal{N}$(0.13, 0.12) & 0.08$\pm$0.09 \rule{0pt}{2.3ex} \rule[-1ex]{0pt}{0pt}\\
\textit{TESS} s40 $c_0$ &  & $\mathcal{U}$(38000, 43000) & 40285$\pm$40 \rule{0pt}{2.3ex} \rule[-1ex]{0pt}{0pt}\\
\textit{TESS} s40 $c_1$ &  & $\mathcal{U}$($-$400, 400) & 2$\pm$12 \rule{0pt}{2.3ex} \rule[-1ex]{0pt}{0pt}\\
\textit{TESS} s40 $c_2$ &  & $\mathcal{U}$($-$600, 600) & 171$\pm$113 \rule{0pt}{2.3ex} \rule[-1ex]{0pt}{0pt}\\
   \textit{GP parameters} & & &  \rule{0pt}{2.3ex} \rule[-1ex]{0pt}{0pt}\\
Rot$_{\rm Q0}$ &  & ... & 0.0010$^{+0.0007}_{-0.0004}$ \rule{0pt}{2.3ex} \rule[-1ex]{0pt}{0pt}\\
Rot$_{\rm deltaQ}$ &  & ... & 0.26$^{+0.10}_{-0.07}$ \rule{0pt}{2.3ex} \rule[-1ex]{0pt}{0pt}\\
Rot$_{\rm fmix}$ &  & ... & 0.0035$^{+0.0013}_{-0.0009}$ \rule{0pt}{2.3ex} \rule[-1ex]{0pt}{0pt}\\
Rot$_\sigma$ s19, s20 &  & ... & 0.0011$\pm$0.0002 \rule{0pt}{2.3ex} \rule[-1ex]{0pt}{0pt}\\
Norm. factor s19, 20, 53, 60 &  & ... & 0.978$\pm$0.001 \rule{0pt}{2.3ex} \rule[-1ex]{0pt}{0pt}\\
Rot$_\sigma$ s26 &  & ... & 0.0012$\pm$0.0002 \rule{0pt}{2.3ex} \rule[-1ex]{0pt}{0pt}\\
Norm. factor s20, 73, 79 &  & ... & 0.977$\pm$0.001 \rule{0pt}{2.3ex} \rule[-1ex]{0pt}{0pt}\\
Rot$_\sigma$ s53 &  & ... & 0.0013$\pm$0.0002 \rule{0pt}{2.3ex} \rule[-1ex]{0pt}{0pt}\\
Rot$_\sigma$ s60 &  & ... & 0.0008$\pm$0.0001 \rule{0pt}{2.3ex} \rule[-1ex]{0pt}{0pt}\\
Rot$_\sigma$ s73 &  & ... & 0.0017$\pm$0.0003 \rule{0pt}{2.3ex} \rule[-1ex]{0pt}{0pt}\\
Rot$_\sigma$ s79 &  & ... & 0.0015$\pm$0.0002 \rule{0pt}{2.3ex} \rule[-1ex]{0pt}{0pt}\\
$V_c$ (RV) & m s$^{-1}$ & $\mathcal{U}$(0, 100) & 4.1$^{+1.5}_{-1.0}$   \rule{0pt}{2.3ex} \rule[-1ex]{0pt}{0pt}\\
$V_r$ (RV) & m s$^{-1}$ & $\mathcal{U}$($-$100, 100) & 11.3$^{+3.1}_{-2.4}$   \rule{0pt}{2.3ex} \rule[-1ex]{0pt}{0pt}\\
$B_c$ (BIS) & m s$^{-1}$ & $\mathcal{U}$($-$50, 50) & 4.0$^{+1.4}_{-0.9}$   \rule{0pt}{2.3ex} \rule[-1ex]{0pt}{0pt}\\
$B_r$ (BIS) & m s$^{-1}$ & $\mathcal{U}$($-$50, 50) & $-$4.0$^{+2.7}_{-3.2}$   \rule{0pt}{2.3ex} \rule[-1ex]{0pt}{0pt}\\
$L_c$ (FWHM) & km s$^{-1}$ & $\mathcal{U}$($-$0.5, 0.5) & 0.013$\pm$0.003   \rule{0pt}{2.3ex} \rule[-1ex]{0pt}{0pt}\\
   \textit{Shared polynomial} & & &  \rule{0pt}{2.3ex} \rule[-1ex]{0pt}{0pt}\\
Poly factor (RV) &  & ... & 0.005$^{+0.012}_{-0.004}$ \rule{0pt}{2.3ex} \rule[-1ex]{0pt}{0pt}\\   
Poly factor (BIS) &  & ... & 0.002$^{+0.008}_{-0.002}$ \rule{0pt}{2.3ex} \rule[-1ex]{0pt}{0pt}\\
Poly factor (FWHM) &  & ... & 0.00001$^{+0.00002}_{-0.00001}$ \rule{0pt}{2.3ex} \rule[-1ex]{0pt}{0pt}\\
Shared $c_2$ &  & ... & $-$0.004$^{+0.005}_{-0.015}$ \rule{0pt}{2.3ex} \rule[-1ex]{0pt}{0pt}\\
Shared $c_3$ &  & ... & $-$0.00003$^{+0.00002}_{-0.00006}$ \rule{0pt}{2.3ex} \rule[-1ex]{0pt}{0pt}\\
\hline
\end{tabular}
\end{table}

\end{appendix}
\end{document}